\documentclass[prl,twocolumn,showpacs,preprintnumbers,amsmath,amssymb]{revtex4}

\usepackage{graphicx}
\usepackage{dcolumn}
\usepackage{bm}

\begin{document}
\title{Doppler cooling with coherent trains of laser pulses and tunable  ``velocity comb''}

\author{Ekaterina Ilinova}
\affiliation{Department of Physics, University of Nevada, Reno,
Nevada 89557, USA}
\author{Mahmoud Ahmad}
\affiliation{Department of Physics, University of Nevada, Reno,
Nevada 89557, USA}
\author{Andrei Derevianko}
\affiliation{Department of Physics, University of Nevada, Reno,
Nevada 89557, USA}

\begin{abstract}
We explore the possibility of decelerating and Doppler cooling an ensemble of two-level atoms by a coherent train of short, non-overlapping laser pulses. We derive analytical expressions
for mechanical force exerted by the train. In frequency space the force pattern reflects the underlying frequency comb structure. The pattern depends strongly on the ratio of the atomic lifetime to the repetition time between the pulses and pulse area. For example, in the limit of short lifetimes, the frequency-space peaks of the optical force wash out. We propose to tune the carrier-envelope offset frequency  to follow the Doppler-shifted detuning as atoms decelerate; this leads to compression of atomic velocity distribution about comb teeth and results in a ``velocity comb'', a series of narrow equidistant peaks in the velocity space.
\end{abstract}

\pacs{37.10.De, 37.10.Gh, 42.50.Wk}

\maketitle
Laser cooling is one of the key techniques of modern atomic physics~\cite{MinLet87Book,MetStr99Book, BerMal10_Book}. Radiative force originates from momentum transfer to atoms from a laser field and subsequent spontaneous emission in random directions. Doppler effect makes the force velocity-dependent.

\begin{figure}[h]
\begin{center}
\includegraphics*[width=2in]{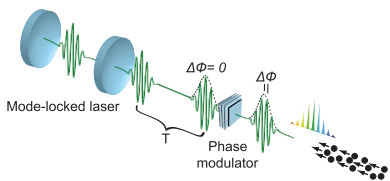}
\end{center}
\caption{(Color online) Schematic of a typical experimental setup.  An atomic beam is slowed and cooled by a train of laser pulses. Phase and shape of pulses may be varied in time to attain optimal cooling.}%
\label{Fig:Setup}%
\end{figure}

Here we develop a systematic theory of Doppler cooling by a {coherent train of short laser pulses} (see Fig.~\ref{Fig:Setup}). A qualitatively new effect comes into play:
atomic quantum-mechanical amplitudes induced by subsequent pulses
{\em interfere} resulting in a periodically varying radiative force as a function of frequency. This structure of the force reflects the comb-like pattern of Fourier image of the pulse train, the so-called frequency comb (FC)~\cite{CunYe03}.
Here we derive the force and show that for sufficiently weak pulses and long atomic lifetimes, each tooth acts as if it were an independent CW laser. In the opposite limit of short lifetimes (short compared to the repetition time between pulses), we recover the force due to an isolated laser pulse. Earlier works on mechanical effects of FCs include proposal involving two-photon transitions~\cite{Kie06}. Following proposal~\cite{Hof88}, pulse trains from mode-locked lasers were also used in cooling experiments~\cite{StrKerKru89,WatOhmTan96}. To the best of our knowledge no analytical analysis of the FC's radiative force has been attempted so far and it is presented here.

Notice that over the past few years the power and spectral coverage of FCs have grown considerably. A fiber-laser-based FC with 10 W average power was demonstrated~\cite{SchHarYos08} and the authors argue that the  technology is scalable above 10 kW average power. The spectral coverage was expanded from optical frequencies to ultraviolet and to IR spectral regions~\cite{AdlCosTho09}. These  advances pave the road for new applications of FCs, such as the laser cooling.

%

As an application, we consider mapping frequency comb to a ``velocity comb''. We demonstrate that during pulse-train cooling, continuous velocity distributions gravitate toward a series of sharp (of a typical Doppler width of $\mathrm{m/s}$ for strong lines and mm/s for weak lines such as intercombination transition in Sr)  peaks in the velocity space. ``Velocity combs'' could be used for studying velocity-dependent (e.g., shape) resonances where traditional beam techniques with their broad velocity distributions would fail~\cite{GilHoeMee06etal}. Moreover, since groups of atoms with different velocities would arrive at the target at different times, the experiment may be carried out ``in parallel'' for many velocities (cf. molecular fingerprinting~\cite{DidHolMbe07}.) Notice that the moniker ``velocity comb'' was used in a work\cite{BanAumSke06} on optical pumping with FCs; we retain this label here as a natural visual for the resulting velocity distribution.

In a typical FC setup, a train of phase-coherent pulses is produced by
multiple reflections of a single pulse injected into an optical
cavity. A short pulse is outcoupled every roundtrip of the
wavepacket inside the cavity, determining a repetition time $T$
between subsequent pulses. At a fixed spatial coordinate, the
electric field of the train may be parameterized as
\begin{equation}
\mathbf{E}(t)=\hat{\varepsilon}\,E_{p}\,\sum\limits_{m}\cos(\omega_{c}%
t-\phi_{m})\,g(t-mT)
\label{Eq:TrainField} \, ,
\end{equation}
where $\hat{\varepsilon}$ is the polarization vector, $E_{p}$ is
the field amplitude, and $\phi_{m}$ is the phase shift. The
frequency $\omega_{c}$ is the carrier frequency
and $g(t)$ is the shape of the pulses. We normalize $g(t)$ so that $\max g( t)
\equiv 1$, then $E_{p}$ has the meaning of the peak amplitude.
While typically pulses have identical shapes and
$\phi_{m}=m~\phi$, one may want to install an active optical
element at the output of the cavity as in Fig.~\ref{Fig:Setup} that could vary the phase and
the shape of the pulses. Also repetition time and intensity of pulses could be controlled by varying reflectivity of cavity mirror.

We focus on two-level systems as these are amendable to analytic treatment and much insight may be gained from analyzing the derived expressions.
Technically, we solve the optical Bloch equations (OBE) for  density
matrix elements (excited and ground state populations are $\rho_{ee}$ and $\rho_{gg}$ and coherences
$\rho_{eg}$ and $\rho_{ge}$)
\begin{eqnarray}
\dot{\rho}_{ee}&=&-\gamma \rho_{ee}+\frac{i}{2}(\rho_{ge} \Omega_{eg}(z,t)
-\rho_{eg} \Omega_{ge}(z,t)),\label{Eq:OBE1}\\
\dot{\rho}_{eg}&=&-(\frac{\gamma}{2}-i\delta_\mathrm{eff})\rho_{eg}+\frac{i}
{2}\Omega_{eg}(z,t)(\rho_{gg}-\rho_{ee}),\label{Eq:OBE2}
\end{eqnarray}
where $\delta_\mathrm{eff}=\delta+\mathbf{k}_c \cdot \bf{v}$ is
the Doppler-shifted detuning ($\delta=\omega_{c}-\omega_{eg}$, $k_c=2\pi/\omega_c$ and $v$ is the
atomic velocity). 
The time- and space- dependent Rabi frequency is
$
\Omega_{ge}(z,t)=\Omega_{p} \, \sum_{m=0}^{N-1} g(t+z/c-m T )e^{i \phi_m} \, ,
$
with the peak Rabi frequency $\Omega_{p}=\frac{E_{p}
}{\hbar}\langle e|\mathbf{D}\cdot\hat{\varepsilon}|g\rangle$  expressed in terms of the dipole matrix element.  Once the OBEs are solved,
radiative force may be determined in terms of the coherence
\begin{equation}\label{ClFrc}
F_z=-p_r\, \mathrm{Im}[\rho_{eg}\Omega_{eg}^*] \, ,
\end{equation}
where $p_r=\hbar k_c$ is the photon recoil momentum.


We start by observing that as long as the duration of the pulse is
much shorter than the repetition time, the atomic system behaves
as if it were a subject to a perturbation by a series of
delta-function-like pulses. In this limit, the only relevant
parameter affecting the quantum-mechanical time evolution is the
effective area of the pulse
$
\theta=\Omega_{p}\,\int\limits_{-\infty}^{\infty}g(t)dt \, .
$
As an illustration we will use a Gaussian-shaped pulse, $g(t)=e^{-t^{2}%
/2\tau_{p}^{2}}$; we will assume that $\tau_p \ll T$.

\begin{figure}[h]
\begin{center}
  \includegraphics*[width=2in]{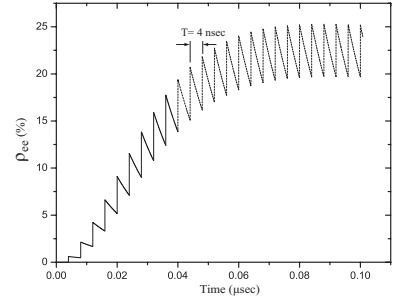}
\end{center}
  \caption
  {Evolution of the excited state population due to interaction with a  train of laser pulses. The atom is initially in the ground state, and it is driven by a train of pulses separated by $T=4 \, \mathrm{ns}$ and of pulse area $\theta=\pi/10$. Radiative lifetime  is $16 \, \mathrm{ns}$.}\label{Fig:Rhoeepls}
\end{figure}

We  distinguish between pre-pulse (left) and post-pulse
(right) elements of the density matrix, e.g., $\left(
\rho_{eg}^{m}\right)  _{l}$ and $\left(  \rho_{eg}^{m}\right)
_{r}$ are the values of coherences just before and just after the
$m^{\text{th}}$ pulse.
%
%
Between the pulses the
dynamics is determined by the spontaneous decay
\begin{align}
\rho_{eg}(t)  &  =\left(  \rho_{eg}^{m}\right)  _{r}\exp\left[
-(\frac
{\gamma}{2}-i\delta_\mathrm{eff}) (t-mT)\right]
,\label{Eq:free}\\
\rho_{ee}(t)  &  =\left(  \rho_{ee}^{m}\right)  _{r}\exp\left[
-\gamma (t-mT)\right]  \,.\nonumber
\end{align}
We neglect the spontaneous decay \emph{during} the
pulse, since for femtosecond pulses, $\tau_p \gamma \ll 1$.
Then
\begin{equation}
\left(  \rho^{m}\right)
_{r}=e^{i~\theta/2~\boldsymbol{\sigma}_{m}}~\left( \rho^{m}\right)
_{l}~e^{-i~\theta/2~\boldsymbol{\sigma}_{m}} \,,
\label{Eq:Rhoaccross}%
\end{equation}
with $\sigma_m = \cos\phi\,\sigma_x-\sin\phi\,\sigma_y$, where $\sigma_{x,y}$ are the Pauli matrices. Analogs of Eqs.~(\ref{Eq:free},\ref{Eq:Rhoaccross}) were derived earlier~\cite{FelBosAci03}.
By stacking  single-pulse (\ref{Eq:Rhoaccross}) and free-evolution (\ref{Eq:free}) propagators, one
may evolve a given initial $\rho$ over duration of the entire
train. In Fig.~\ref{Fig:Rhoeepls} we show results of such calculation
for the excited state population (atom remains at rest).

Now we focus on the evaluation of the
radiative force. The laser field is present only during the
pulse, so we deal with a sum over instantaneous forces.
The change in the atomic momentum  due to a single pulse is
\begin{equation}
\frac{-\Delta\mathbf{p}_{m}}{p_{r}}=\left(  \left(
\rho_{ee}^{m}\right) _{r}-\left(  \rho_{ee}^{m}\right)
_{l}\right)  ~\mathbf{\hat{k}}_{c} \, ,
\label{Eq:dPm}%
\end{equation}
i.e., a  laser pulse
imparts a fractional momentum kick equal to the difference of
populations before and after the pulse. Since $0\leq\rho_{ee}\leq1$, the maximum momentum kick per pulse is equal to the recoil momentum.

By combining Eqs.~(\ref{Eq:free},\ref{Eq:Rhoaccross},\ref{Eq:dPm})  we find the radiative force.  Time evolution of population, Fig.~\ref{Fig:Rhoeepls},  separates into two regimes: initial transient phase and the quasi steady-state (QSS) regime when radiative-decay-induced drop in the population following a given pulse is fully restored by the subsequent pulse. Doppler cooling requires many scattering cycles and we focus on the QSS (or ``the coherent accumulation''~\cite{FelBosAci03}) regime.

In the QSS regime,  $\rho_{ee}\left(  t\right) =\rho_{ee}\left(
t+nT\right)$ and pre- and post-pulse values
$\left(  \rho_{ee}^{m}\right)_{l,r}$ do not depend on the pulse
number $m$; we simply denote these values as $\left(
\rho_{ee}^{s}\right)
_{l,r}$. Then  Eq.(\ref{Eq:dPm}) becomes $
-\Delta p_{s}/p_{r}=\left(  \rho_{ee}^{s}\right)
_{r}~\times\left(
1-e^{-\gamma T}\right) 
$.
We find $\left(\rho_{ee}^{s}\right)$ using non-perturbative propagators Eqs.~(\ref{Eq:free},\ref{Eq:Rhoaccross}) and arrive at the fractional momentum kick per pulse
\begin{equation}
\frac{-\Delta\mathbf{p}_{s}}{p_{r}}=\frac{
\sin^{2}\left(  \theta/2\right)  \sinh\left(  \gamma T/2\right)  }{\cosh\left(
\gamma T/2\right)  - \cos^{2}(\theta/2)\,\cos \eta }\mathbf{\hat{k}}_{c} \,.\label{Eq:dPmSaturationExact}%
\end{equation}
Here the Doppler-shifted phase $\eta$ is
\begin{equation}
\eta=(\delta +\mathbf{k}_{c}\cdot\mathbf{v})T-\phi.
\label{Eq:phaseDef}%
\end{equation}
Finally, the radiative force is $F_\mathrm{train}= \Delta p_s / T$.


\begin{figure}[h]
\begin{center}
\includegraphics*[width=3.2in]{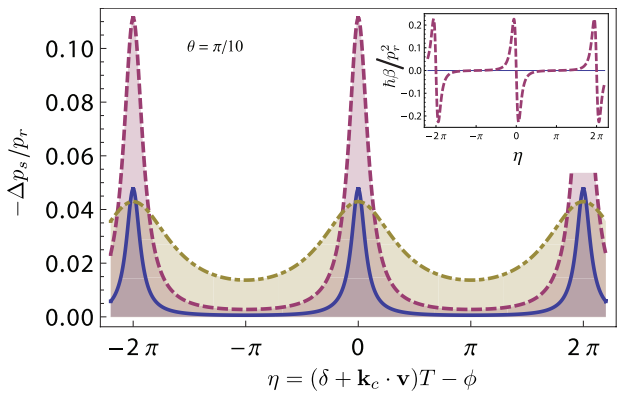}
\end{center}
\caption{(Color online) Frequency dependence of the fractional
momentum for a pulse area $\theta=\pi/10$. The three curves differ by the values of the
parameter $\gamma T$ (solid curve $\gamma T=0.1$, dashed $\gamma T=0.447$ and dot-dashed $\gamma T=2.5$). The spikes
reflect the underlying frequency-comb teeth structure of the
pulse train in the frequency domain.
Inset: friction coefficient $\beta$ as a function of phase for the dashed curve of the main panel.}%
\label{Fig:SatThetaOver10}%
\end{figure}

As a function of phase $\eta$ (or frequency or velocity),
the force spikes at the positions of the Doppler-shifted frequency comb teeth,
$\eta_{n}=2\pi\times n$, with $n$ being integer numbers (see  Fig.~\ref{Fig:SatThetaOver10}.
In Fig.~\ref{Fig:SatThetaOver10} we also investigate
dependence on the values of  parameter $\gamma T$.
Let us focus on one of the teeth (e.g.,
$\eta=0$). As $\gamma T$ is increased the momentum kick grows,
reaches maximum, and then declines; apparently for a given
$\theta$ there is
an optimal value of $\gamma T$. By analyzing Eq.~(\ref{Eq:dPmSaturationExact}%
), we find this optimal value to be
$
\left(  \gamma T\right)_{\mathrm{opt}}^F=2\cosh^{-1}\left(
1/\cos^{2}\left(  \theta/2\right) \right).
$
For example, for $\theta=\pi/10$, the optimal value is $\left(
\gamma T\right) _{\mathrm{opt}}^F \approx0.447$, i.e., the
radiative lifetime is roughly twice the repetition period.

Eq.~(\ref{Eq:dPmSaturationExact}) is non-perturbative.
It remains valid even for strong laser pulses, as long as the pulses do not overlap. For $\theta = 2 \pi$ the force vanishes since $2 \pi$-pulse does not redistribute population.
For $\pi$-pulses, the r.h.s.\  of Eq.~(\ref{Eq:dPmSaturationExact}) reduces to a
frequency-\emph{independent} value $\tanh(\gamma T/2)$.

In the limiting case of weak pulses ($\theta
\ll 1$) and fast repetition rates ($\gamma T\ll1$), Eq.~(\ref{Eq:dPmSaturationExact}) reduces to
\begin{eqnarray}
\lefteqn{\left. F_\mathrm{train} \right|_{\gamma T \ll 1, \theta \ll 1} \approx p_{r}\Omega_{p}^{2}\left(
\frac{\sqrt{\pi}\tau_{p}}{T}\right)^2\,\times} \nonumber\\
&&\sum_n
\frac{\gamma/2}{\left(  \gamma/2\right)  ^{2}+\left(  \delta
+\mathbf{k}_{c}\cdot\mathbf{v}_{0}-\left(  \phi+2\pi
n\right) /T\right)^{2}} \, . \label{Eq:IndepLasersForce}
\end{eqnarray}
By comparing
Eq.~(\ref{Eq:IndepLasersForce}) with relevant CW expressions (see, e.g., Ref.~\cite{BerMal10_Book}), we arrive at the qualitative
picture where each tooth acts as an independent CW laser,
intensity of which has been reduced by the factor of
$\sqrt{\pi}\tau_{p}/T$. This factor is roughly equal to the number
of  teeth fitting inside the overall frequency envelope of the FC.


Does a single tooth have enough power for cooling?
The relevant parameter is the saturation intensity, $I_s$. Its typical value (sodium atom) is $6.4~\mathrm{mW/cm^{2}}$. Typical FC parameters ($\tau_p=100 \, \mathrm{fs}$, $T = 1 \, \mathrm{ns}$ and  average power $1 \, \mathrm{W}$) translate into the power per tooth of $0.1 \, \mathrm{mW}$, thereby $I_s$ can be attained by focusing the laser output to a spot of 1.4 mm diameter. Notice that a fiber-laser-based FC with 10 W average power has been demonstrated~\cite{SchHarYos08} and the authors argue that the demonstrated technology is scalable to 10 kW average power. With this new generation of combs the cross-section of the interaction region may be increased dramatically.

Now we turn to the dynamics of slowing down and cooling an entire atomic ensemble,
characterized by some velocity distribution $f(v,t)$ (time-dependence is caused by radiative force).
To be specific, consider a typical use of radiative force for slowing down an atomic beam. The laser pulses  would impinge on  the atoms (see Fig.~\ref{Fig:Setup}) countering their motion. The radiative force (\ref{Eq:dPmSaturationExact}) depends on  the atomic
velocity via Doppler shift. As velocity is varied across the ensemble, the maxima of the force
 occur at discrete values of velocities ($n$ are integers)
\begin{equation}
v_n = ( 2 \pi \, n - T \,\delta + \phi)/(k_c T) . \label{Eq:VelocityTeeth}
\end{equation}
The force peaks are separated by $v_{n+1}-v_n = \lambda_c/T$ in the velocity space.
The comb may have multiple teeth effectively interacting with the ensemble.


Cooling can be characterized by introducing friction coefficient $\beta$,
$
  F_\mathrm{train}(v + \Delta v ) \approx F_\mathrm{train}(v) - \beta(v) \Delta v  \, .
  \label{Eq:ForceExpansionBeta}
$
If $\beta > 0$, there is a compression of velocity distribution around $v$.
In the limiting case of $\gamma T \gg 1$ or $\theta =\pi$ the force does not depend on velocity,
thereby $\beta =0$ and while the ensemble slows down, there is no cooling.
%
The friction coefficient  may be derived analytically from the force
(\ref{Eq:dPmSaturationExact}).
We plot the dependence of $\beta$ on $\eta$ in Fig.~\ref{Fig:SatThetaOver10}.
It acquires the maximum
value at $\eta = \bar{\eta}$,
\begin{eqnarray}
\lefteqn{\cos \bar{\eta}=\frac{1}{2} \sec
^2\left({\theta }/{2}\right) \times } \label{etaz} \\
&&\left(\sqrt{8 \cos ^4\left({\theta
}/{2}\right)+\cosh ^2(\gamma T /2 )}-\cosh (\gamma T/2 )\right) \, .
\nonumber
\end{eqnarray}
In the CW limit, this expression leads to detuning of $\gamma/2$ below the atomic resonance as expected. One could optimize $\beta$ by varying $\theta$ or $\gamma T$. 

As the atoms slow down, they come in and out of resonances with the FC teeth, leading to periodic variation in the sign of $\beta$; no cooling results due to this variation. To keep FC teeth in resonance with  the Doppler-shifted atomic transition, one may vary the offset phase $\phi$ by an optical element installed at the cavity output
(see Fig.~\ref{Fig:Setup} and Eq.~(\ref{Eq:VelocityTeeth})).
If for a given velocity group  initially centered  at $v_{mp}(t=0)$, the phase detuning is kept at $\bar{\eta}$, there will be a compression of velocity distribution around $v_{mp}(t)$.
We may satisfy this requirement by tuning the phase according to $\phi(t) = \left( \delta + k_c v_{mp}(t) \right) T - \bar{\eta}$.  As $v_{mp}(t)$ becomes smaller
due to radiative force, the offset phase needs to be reduced.
Using Eq.(\ref{Eq:dPmSaturationExact}) we  find the required pulse-to-pulse decrement of the phase
\begin{equation}
\Delta \phi_T=\frac{p_r^2 T}{\hbar M_a}\frac{\sin^{2}\left( \theta/2\right)  \sinh\left( {\gamma T}/{2}\right)  }{
\cos\bar{\eta}\cos^{2}\left({\theta}/{2}\right)-\cosh\left( {\gamma T}/{2}\right)
}\,.\label{Eq:DeltaPhi}%
\end{equation}

We would like to emphasize an important distinction between  earlier works~\cite{StrKerKru89,WatOhmTan96,Hof88} and our approach. This is
related to  the difference between slowing and cooling. In~\cite{StrKerKru89,WatOhmTan96,Hof88}  there is no substantial compression of velocity distribution about the frequency teeth. One could easily see from figures in~\cite{StrKerKru89} that the spread of velocities about a single tooth is
approximately equal to half the distance between  neighboring teeth. There
is no velocity compression (cooling) about individual teeth because no frequency tuning was done in those
papers. Linguistically ``velocity comb'' implies {\em narrow} teeth in the velocity
space. Eq.~(\ref{Eq:DeltaPhi}) prescribes how to achieve this narrowing via phase tuning. Each tooth
ultimately would have width in the order of Doppler width or smaller.

When the phase offset is driven according to (\ref{Eq:DeltaPhi}),
the entire frequency-comb structure shifts towards lower frequencies.
As the teeth sweep through the velocity space, atomic $v(t)$ trajectories are ``snow-plowed''
by teeth, ultimately leading to narrow velocity spikes collected on the teeth.
Formally, we may separate initial velocities into groups
$v_{mp}(t=0)+\left(\bar{\eta} + 2\pi n \right)/k_c T
< v(t=0) < v_{mp}(t=0)+\left( \bar{\eta} + 2\pi (n+1) \right)/k_c T,  n=0,\pm 1 \ldots$
The width of each velocity group is equal to the distance between neighboring teeth in
velocity space, $\lambda_c/ T$.  As a result of ``snow-plowing'', the $n^\mathrm{th}$ group
will be piled up at $v_n(t)=v_{mp}(t)+  n \lambda_c/ T$. The final velocity spread of individual
velocity groups will be limited by the Doppler temperature, $T_D = \hbar \gamma/2 k_B$.

\begin{figure}[h]
\begin{center}
\includegraphics*[width=3.2in]{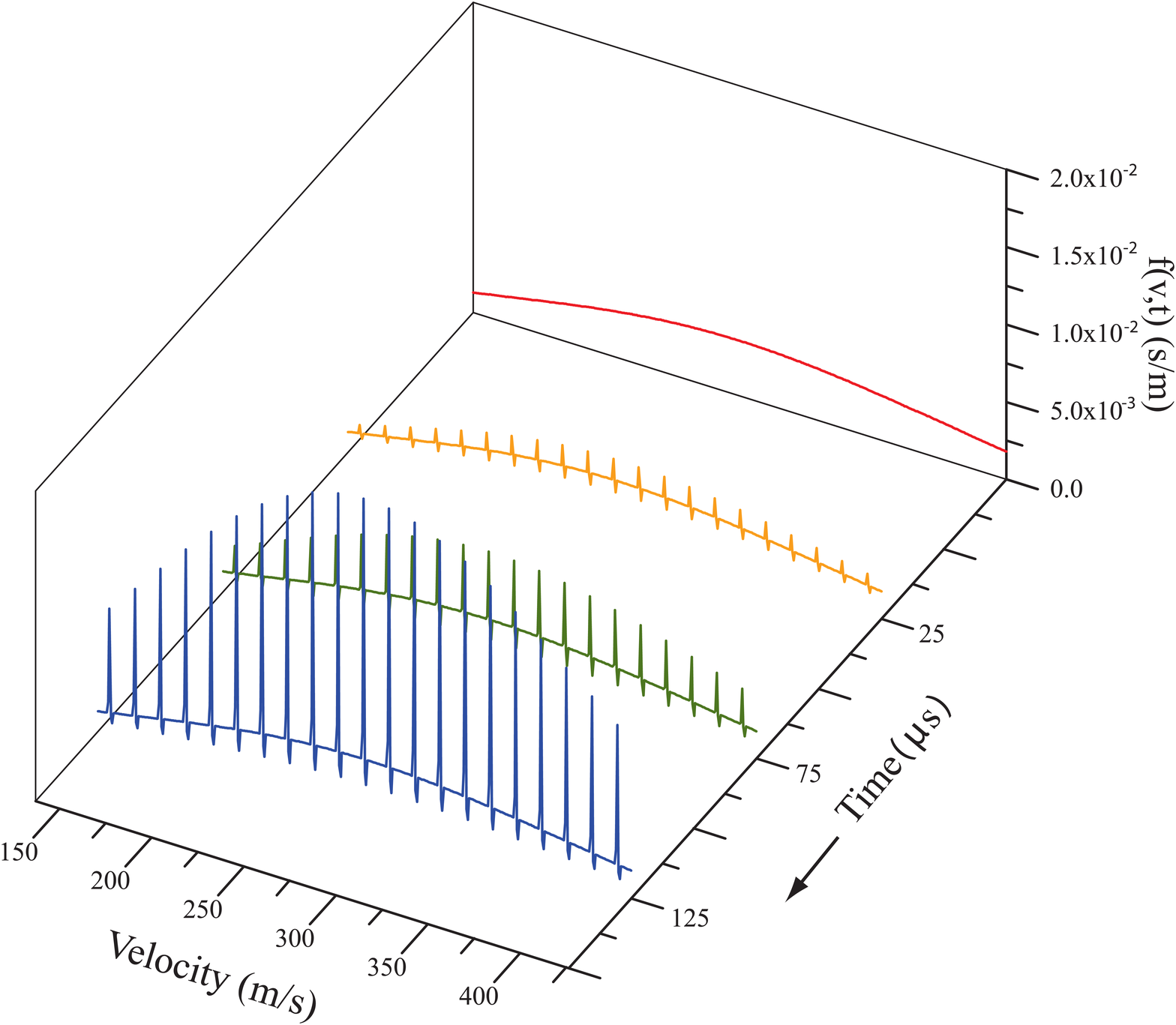}
\end{center}
\caption{Time-evolution of velocity distribution for a thermal beam subjected to
a coherent train of laser pulses. Pulse-to-pulse phase offset of the train is varied
linearly in time  as prescribed by Eq.~(\ref{Eq:DeltaPhi}). Atomic and pulse train parameters are:
$\gamma T=0.0026$, $\theta=0.0019$. The optimal phase detuning is $\bar{\eta}=0.001$.
Characteristic initial temperature of the ensemble is $293 \, \mathrm{K}$.}
\label{Fig:fvt}%
\end{figure}

The formation of velocity comb is illustrated in Fig.~\ref{Fig:fvt} where we
consider cooling and slowing a 1D thermal beam of $^{88}$Sr by a pulse train.  The initial velocity distribution is characterized by $f(v, t=0)=\frac{9}{2} \frac{v^3}{v_{mp}^4}\exp(-\frac{3v^2}{2 v_{mp}^2})$,
where $v_{mp}$ is the most probable velocity at $t=0$. In this example we use the weak $5s^2\,^1\!S_0 \rightarrow 5s5p\,^3\!P_1$ transition with
$\gamma=5.3\times 10^{4}\, 1/\mathrm{s}$. Parameters of the train are $T=50 \,\mathrm{ns}$ and $\theta=0.0019$. At the end of the process we end up with velocity comb separated by 13.8 m/s and of Doppler-limited width of 7.6 mm/s (this is comparable to the recoil limit). About 14\% of the total number
of atoms is ``snow-plowed'' into the teeth within 125 $\mu\mathrm{s}$.

 Notice that by shining two counter-propagating pulse trains on the atoms, one could control $v$-positions of velocity teeth at will, as shifting phase of one train with respect to the other would change the balance of two counter-acting radiative forces exerted by the trains.


We demonstrated that radiative force exerted by laser pulse trains has unique features and expands the toolbox of laser cooling techniques.  For example, one may engineer velocity combs that may be used for studies of narrow collision resonances and thresholds \cite{GilHoeMee06etal,NiOka10}.  In some cases, the frequency comb may be already a part of experimental setup, e.g., in optical atomic clocks \cite{Hal06}. By using it for cooling would reduce the number of lasers. Also the setup does not require Zeeman slowers, whose fields may be detrimental for precision measurements~\cite{ZhuOatHal91}.

We would like to thank D. Budker, M. Gruebele, M. Kozlov, E. Luc-Koenig, and J. Weinstein for discussions. This work was supported in part by the NSF.


\end{document}